\documentclass[]{spie}  %>>> use for US letter paper
 \usepackage{comment}
\usepackage{amsmath,amsfonts,amssymb}
\usepackage{graphicx}
\usepackage[colorlinks=true, allcolors=blue]{hyperref}

\title{Development of dual-polarization LEKIDs for CMB observations}

\author[a]{Heather McCarrick}
\author[a]{Maximilian H. Abitbol}
\author[e]{Peter A.R. Ade}
\author[e]{Peter Barry}
\author[b]{Sean Bryan}
\author[b]{George Che}
\author[c]{Peter Day}
\author[e]{Simon Doyle}
\author[a]{Daniel Flanigan}
\author[a]{Bradley R. Johnson}
\author[a]{Glenn Jones}
\author[c]{Henry G. LeDuc}
\author[a]{Michele Limon}
\author[b]{Philip Mauskopf}
\author[a]{Amber Miller}
\author[e]{Carole Tucker}
\author[c,d]{Jonas Zmuidzinas}

\affil[a]{Department of Physics, Columbia University, New York, NY 10025, USA}
\affil[b]{School of Earth and Space Exploration, Arizona State University, Tempe, AZ 85287, USA}
\affil[c]{Jet Propulsion Laboratory, Pasadena, CA 91109, USA }
\affil[d]{Caltech, Pasadena, CA 91109, USA}
\affil[e]{School of Physics and Astronomy, Cardiff University, Cardiff, Wales CF24 3AA, UK}

\authorinfo{H.M.: E-mail: hlm2124@columbia.edu}

\begin{document} 
\maketitle

\begin{abstract}
We discuss the design considerations and initial measurements from arrays of dual-polarization, lumped-element kinetic inductance detectors (LEKIDs) nominally designed for cosmic microwave background (CMB) studies. The detectors are horn-coupled, and each array element contains two single-polarization LEKIDs, which are made from thin-film aluminum and optimized for a single spectral band centered on 150 GHz. We are developing two array architectures, one based on 160 micron thick silicon wafers and the other based on silicon-on-insulator (SOI) wafers with a 30 micron thick device layer. The 20-element test arrays (40 LEKIDs) are characterized with both a linearly-polarized electronic millimeter wave source and a thermal source. We present initial measurements including the noise spectra, noise-equivalent temperature, and responsivity. We discuss future testing and further design optimizations to be implemented.\end{abstract}


%%%%%%%%%%%%%%%%%%%%%%%%%%%%%%%%%%%%%%%%%%%%%%%%%%%%%%%%%%%%%%%%%%%%%%%%%%%%%%%%%%%%%%%%%%%%%%%
\section{INTRODUCTION}
\label{sec:intro}  % \label{} allows reference to this section

Lumped element kinetic inductance detectors (LEKIDs) are superconducting resonators, which are also photon detectors.
The resonance is set by a capacitor and inductor, the latter of which has both geometric and kinetic components. 
The LEKID inductor acts as the absorber and is impedance matched to the incoming radiation. 
The absorbed photons break Cooper pairs which changes the quasiparticle density and subsequently the surface impedance and kinetic inductance.  
This results in a shift in resonance frequency and quality factor both of which are read out through a transmission line. 
LEKIDs, and more broadly microwave kinetic inductance detectors (MKIDs), have been successfully demonstrated and deployed for a range of frequencies\cite{Mazin2013,nikatsz13}. 
Photon-noise limited performance has been shown in multiple frequency bands as well\cite{mauskopf14, Hubmayr2014}.
For cosmic microwave background (CMB) polarization studies it is important that the detector noise to be sub-dominant to the photon noise. 
Current CMB experiments employ thousands of detectors. 
To further increase the sensitivity, it is necessary to increase the pixel count. 
LEKIDs are a natural candidate as hundreds of detectors\cite{McHugh2012, baselmans2015} can be read out on a single transmission line.
We are developing dual-polarization LEKIDs that have two resonators within a single optical element. 
The two resonators are sensitive to orthogonal polarizations for observation at a frequency band centered at 150~GHz. 
Dual-polarization LEKIDs will effectively double the number of detectors for a given focal plane area compared to single polarization detectors. 
LEKIDs have been demonstrated as sensitive devices for absorbing single-polarization radiation at millimeter wavelengths\cite{flanigan_2016}, and dual-polarization radiation for far infrared\cite{Dober_2016}.
In this proceedings we present (i)  design considerations for dual-polarization LEKIDs at millimeter-wavelengths, (ii) initial test results, and (iii) steps for further optimization and testing. 
%

%%%%%%%%%%%%%%%%%%%%%%%%%%%%%%%%%%%%%%%%%%%%%%%%%%%%
 \begin{figure} [t!]
   \centering
   %\begin{center}
   \begin{tabular}{c} %% tabular useful for creating an array of images 
   \includegraphics[height=6.6cm]{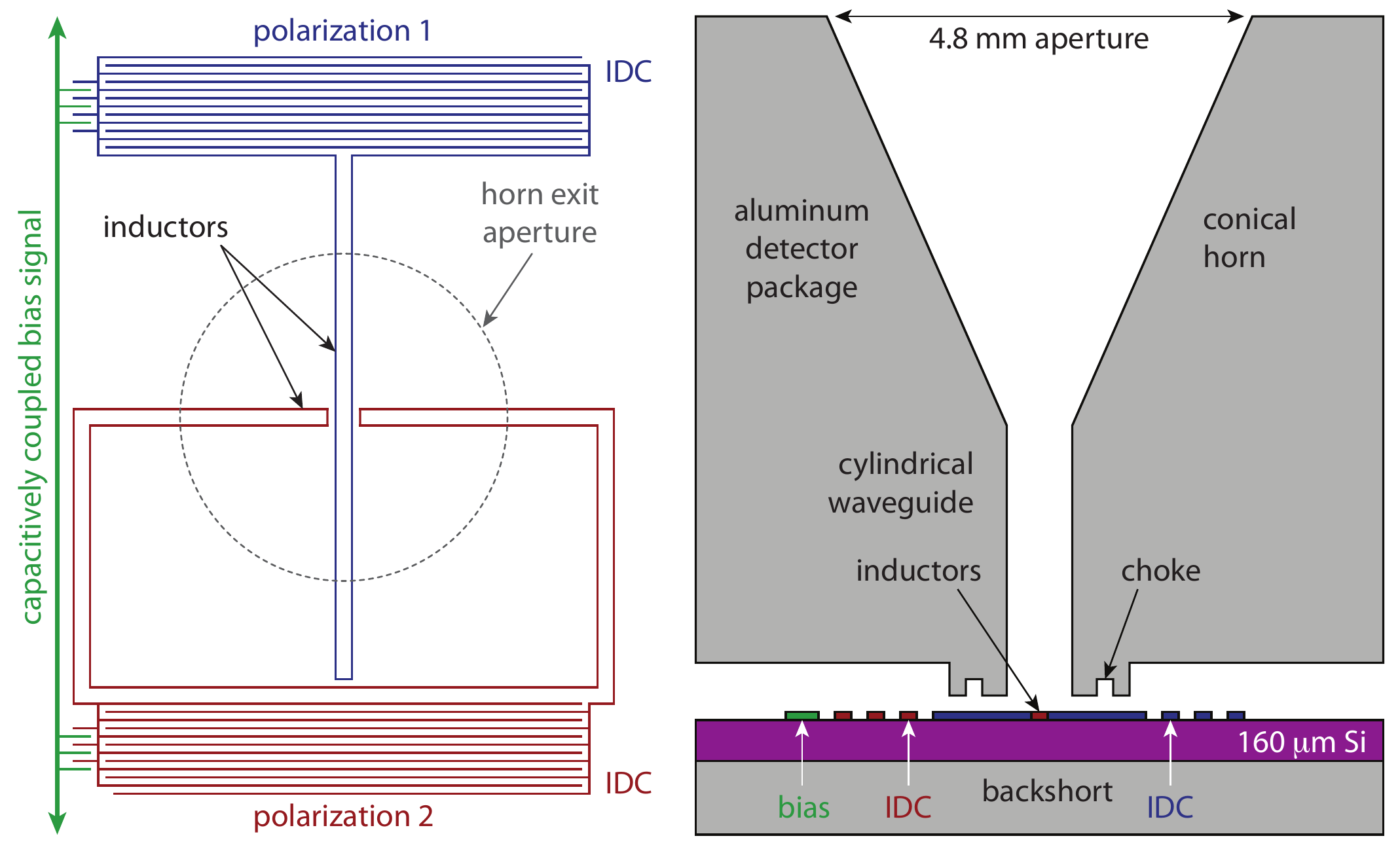}    
   \includegraphics[height=6.5cm]{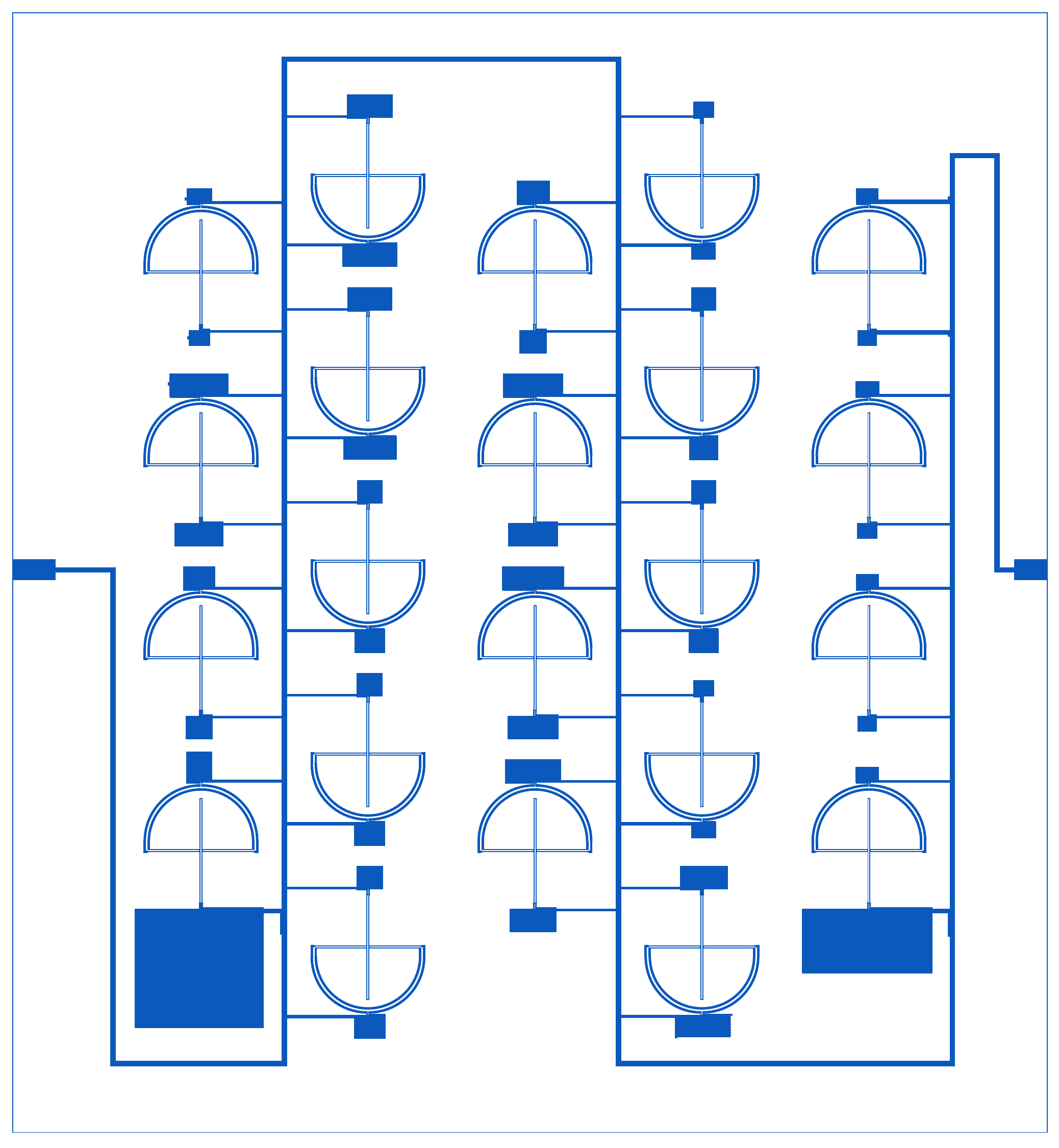}
   \end{tabular}
   %\end{center}
   \caption[example] 
   { \label{fig:detector_schematic}
{\sl Left:} Schematic of a single detector. The resonators corresponding to the orthogonal polarizations are shown in red and blue. The dotted circle represents the waveguide exit aperture. The resonator inductor acts as the photon absorber. The resonators are capactively coupled to the transmission line. 
{\sl Center:} Cross-section view of a single-element. The horn aperture tapers to a cylindrical waveguide which also acts as a high-pass filter. A choke impedance matches between the waveguide and device while also controlling lateral radiation loss. The detectors are fabricated on silicon and directly illuminated. The package bottom acts as the backshort, the distance of which is set by the silicon wafer thickness.
{\sl Right:} Layout of the test array. There are 20-elements, or 40 resonators, that are horn coupled and 2 dark elements. A single transmission line reads out all the devices. Each resonator has a unique resonance frequency set by the IDC value.}
\end{figure} 

%%%%%%%%%%%%%%%%%%%%%%%%%%%%%%%%%%%%%%%%%%%%%%%%%%%%%%%%%%%%%%%%%%%%%%%%%%%%%%%%%%%%%%%%%%%%%%%
\section{Design Considerations}
\label{sec:design}

\subsection{Design Overview}
The dual-polarization LEKID design is shown in Fig.~\ref{fig:detector_schematic}. 
The resonators consist of two orthogonal inductors connected to interdigitated capacitors (IDC) and are coupled capactively to the transmission line. 
The focal plane architecture is heavily based on our experience with single-polarization KIDs\cite{mccarrick_2014}.
A conical horn with a 4.8~mm aperture narrows to a cylindrical waveguide, which acts as a high-pass filter. 
A choke is used to control crosstalk and helps impedance match to the detector. 
The LEKIDs are horn coupled. 
The devices are fabricated on high-resistivity silicon.  
The wafer thickness of  $\lambda / 4$ sets the distance to the backshort. 
The detector package acts as the backshort, increasing optical efficiency. 
The layout of a 20-element test array is shown in Fig.~\ref{fig:detector_schematic}. 
Each element consists of two resonators as described above, and the elements are arranged in a 4.8~mm hexagonal pitch.  
There are 44-resonators for each prototype array, with 2 dark elements and 20 coupled to conical horns. 
The 20-element arrays are prototypes for a 271-element close-packed hex array with 542 detectors. 

\subsection{Design Requirements}
\label{sec:designrequirements}
\begin{figure} [t]
   \centering
   %\begin{center}
   \begin{tabular}{c} 
   \includegraphics[height=7.1cm]{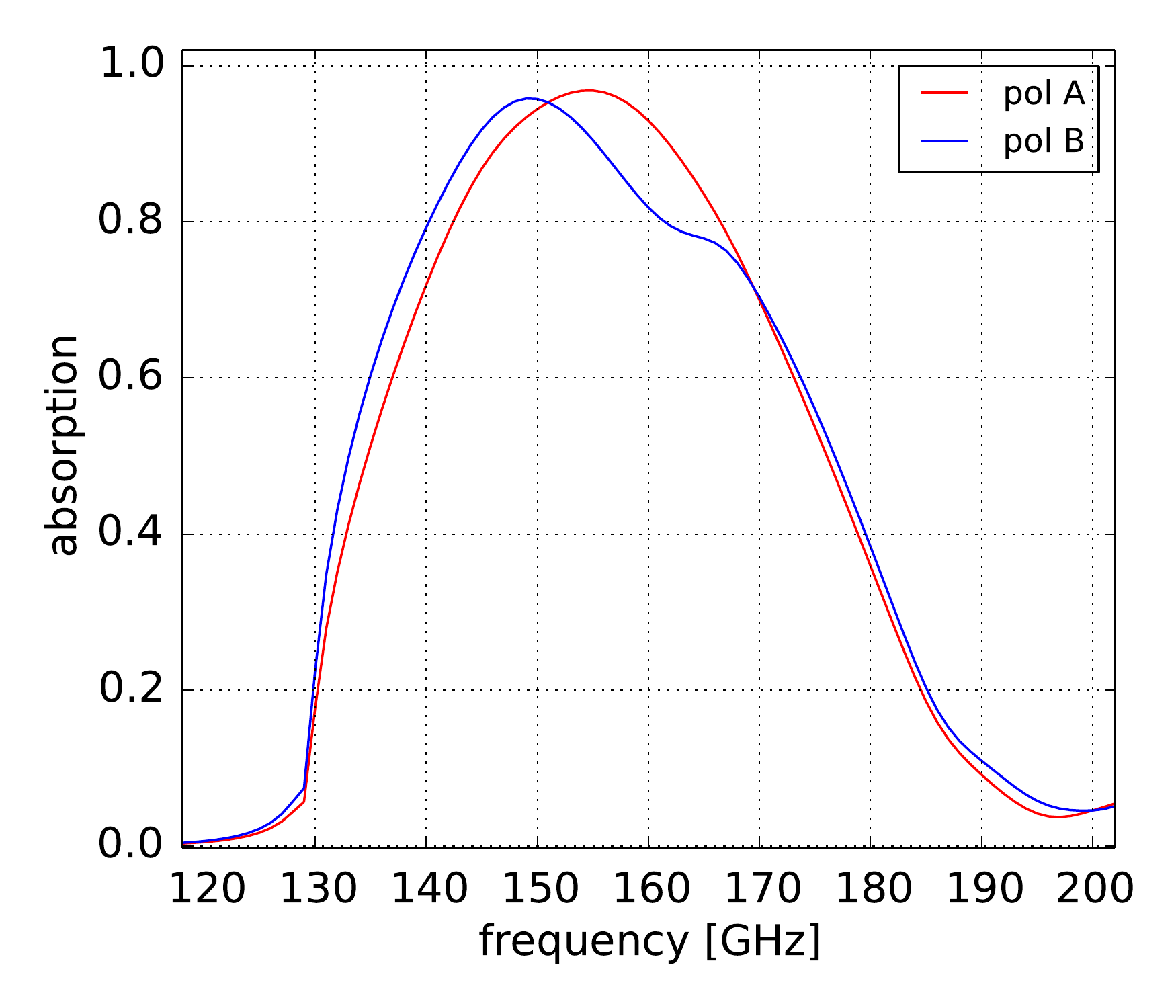}    
   \includegraphics[height=7.1cm]{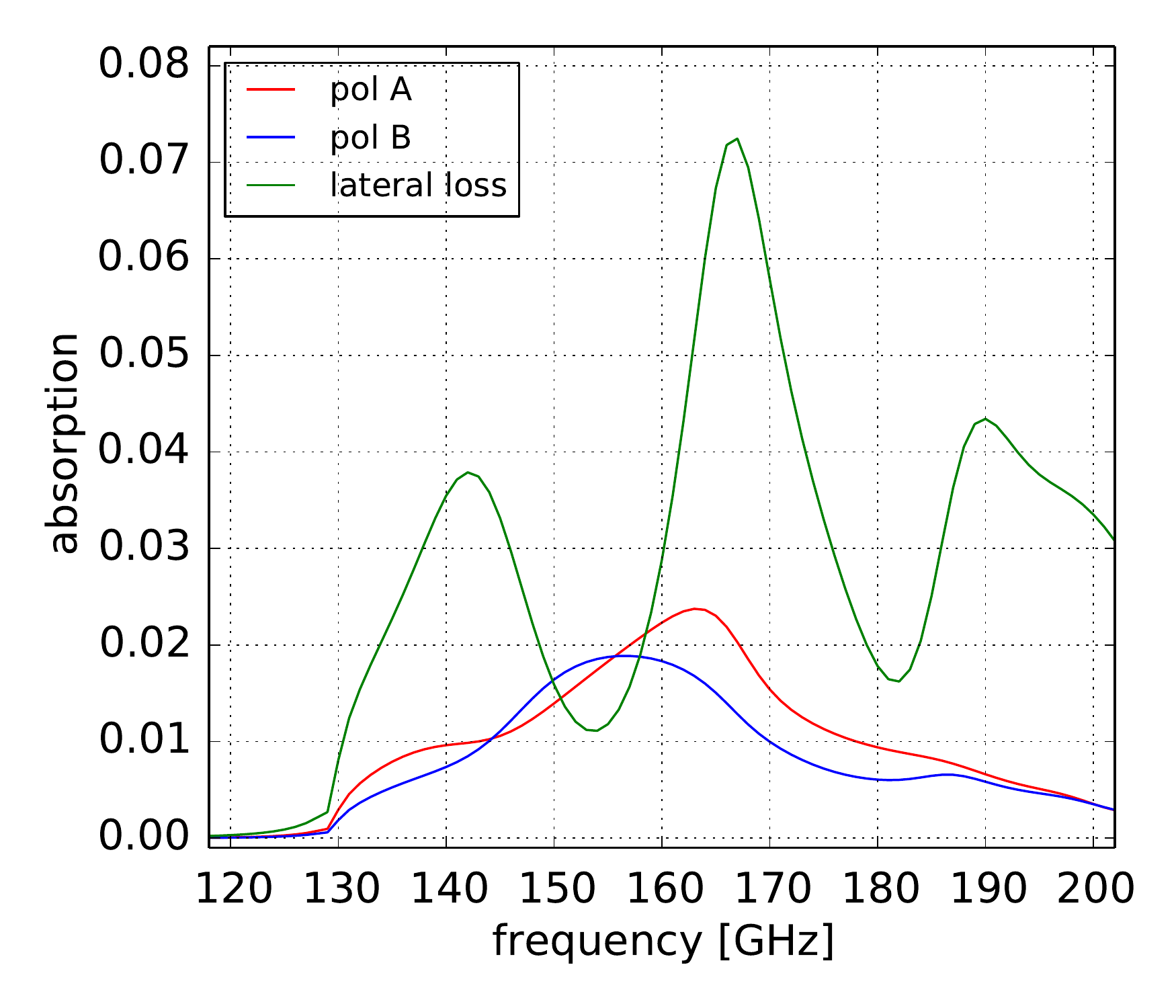}
   \end{tabular}
   %\end{center}
   \caption[example] 
   {\label{fig:hfss} 
	{\sl Left:} HFSS simulations of the front-illuminated devices show the absorption spectra for both polarizations. A waveguide defines the high-pass and a metal mesh filter defines the low-pass. 
	{\sl Right:} HFSS simulation showing the cross-polarization absorption and the lateral loss, which can be interpreted as the maximum possible detector-to-detector optical crosstalk. This HFSS simulation reflects the design show in Fig.~\ref{fig:detector_schematic}. As built and tested for the prototype 20-element arrays, the gap between the detectors and waveguide was slightly larger and the chokes not included. This has the effect of decreasing the optical efficiency and increasing the lateral loss by a few percentage points. 
}
 \end{figure}

The detectors have many design parameters which must be optimized to achieve maximal performance. 
The optimal design parameters for different aspects of detector performance are often in competition.
Here we discuss the important requirements considered. 
The optical coupling is controlled both by the absorber design and the rest of the optical elements, from horn to backshort, or the focal plane design. 
The optimization of the absorber is similar to that as described in Bryan, et al\cite{bryan_2015a}.
In the original design, the LEKIDs were deposited on silicon-on-insulator (SOI)  and coupled to through the silicon, much of which is removed from behind the device through a DRIE-process. 
This approach is beneficial because it offers high-optical efficiency over a wide bandwidth. 
However, this approach involves a complex fabrication process, and we are still developing this architecture.
A simpler approach is to use the wafer thickness itself to set the backshort distance. 
For this architecture, LEKIDs of a similar absorber design as in the SOI-architecture are deposited on silicon. 
The array is mounted on the package bottom which acts as the backshort. 
The absorbers are long traces.
The inductors for the orthogonal polarizations lay perpendicular to one another. 
The inductors are naturally polarization sensitive, preferentially absorbing radiation with the E-field aligned to the thin inductor trace. 
A plot showing the simulated absorption efficiencies for the two polarizations and the cross-polarization response is shown in Fig.~\ref{fig:hfss}. 
The crosstalk between detectors is controlled in the following ways.
The area of the absorbing element is small, as compared to many meandered inductors used in other LEKID designs. 
The radiation can propagate laterally\cite{mccarrick_2015} and we minimize the opportunity to do so by decreasing the gap between the detectors and waveguide exit aperture. 
Additionally, the waveguide choke suppresses the laterally propagating leakage radiation, as well as improving the impedance match to the absorbers.  
The simulated maximum crosstalk is shown in Fig.~\ref{fig:hfss}. 
The readout frequency is centered around 1~GHz.
We targeted 1~GHz because of the space available for the capacitor given the 4.8~mm pitch and the availability of microwave mixers in that frequency range. 
Our FPGA-based readout supports a bandwidth of 500~MHz.
In order to achieve a resonance frequency spacing of 10 times the resonance width to avoid collisions between 542 detectors, the quality factors $Q$ of the resonator needs to be $>~10,000$. 
The detector noise is made up of contributions from generation-recombination (GR), two-level systems (TLS), and amplifier noise. 
We designed the detectors to minimize TLS noise by using relatively large gaps between the capacitor plates and targeting a low resonance frequency.
The amplifier noise can be lowered below the device noise by adjusting the readout power up until the point that non-linear effects come into play. 
Many of the factors controlling detector noise and responsivity are set above by the optical requirements for the device. 
The noise mechanisms are quantitatively described more fully in the literature\cite{zmu,mccarrick_2014,flanigan_2016}.
%

%%%%%%%%%%%%%%%%%%%%%%%%%%%%%%%%%%%%%%%%%%%%%%%%%%%%%%%%%%%%%%%%%%%%%%%%%%%%%%%%%%%%%%%%%%%%%%%

\section{Initial Testing}
\label{sec:sections}
\begin{figure} [t]
   \centering
   %\begin{center}
   \begin{tabular}{c} 
   \includegraphics[height=7.6cm]{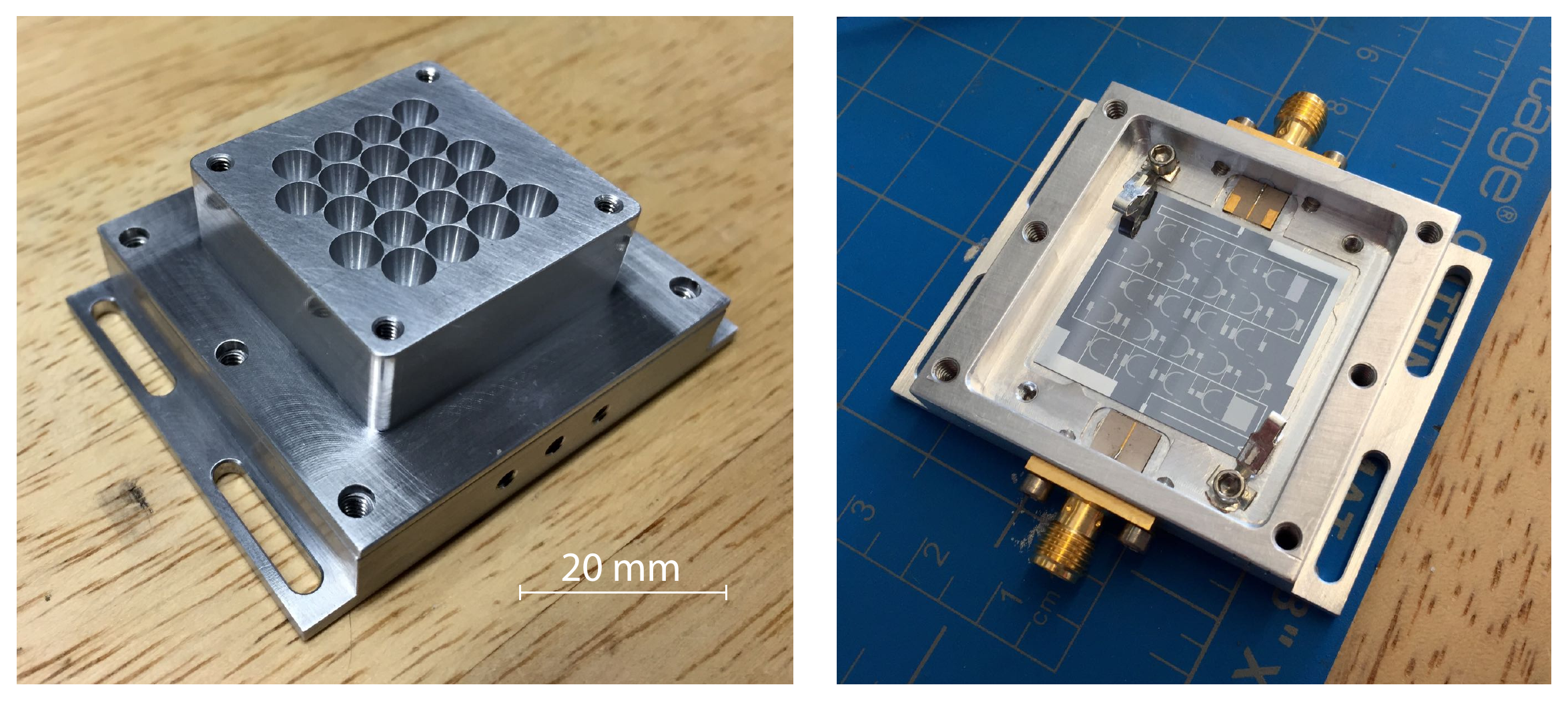}
   \end{tabular}
   %\end{center}
   \caption[example] 
   { \label{fig:package} 
{\sl Left:} Photograph of the dual-polarization test module. This module contains 44~dual-polarization LEKIDs. The conical horns (visible in photo) taper to a waveguide that acts as a high-pass filter. At the exit of the waveguide is a choke to control crosstalk. The incoming radiation is then absorbed by the detector, behind which is a $\sim \lambda$/4 backshort. 
{\sl Right:} Photograph of a dual-polarization test LEKID chip in the aluminum test package with the horns removed.}
 \end{figure}

The LEKIDs were tested in an adiabatic demagnetization refrigerator (ADR) cryostat. 
The detector package, pictured in Fig.~\ref{fig:package}, is mounted on a 100~mK cold stage. 
A low-pass quasi-optical metal mesh filter sits on top of the horns.
The high-pass filter is provided by the cylindrical waveguides.  
A variable blackbody load illuminates the detectors.
The blackbody can be regulated between 2~-~7~K. 
Additionally, an electronic millimeter-wave source can be swept from 140~-~165~GHz.
The tests are similar to those in papers previously published by this collaboration\cite{mccarrick_2014, flanigan_2016}, where they are described in more detail.
Dark tests, with the horns covered, were performed at JPL.

The internal quality factors of the devices measured dark were measured to be $\sim10^6$.  
The coupling quality factors therefore predominantly set the resonator quality factor, and were in the $10^4$ range.  
When loaded by a beam-filling 3~K blackbody, the quality factors were approximately $10^4$. 
The quality factors are sufficient for the multiplexing requirements at these blackbody temperatures. 
The yield for this test array was excellent: 100$\%$ of the resonators were present. 
For a ground based experiment, we expect a higher loading, which would appreciably degrade the quality factor of these resonators.
As discussed below, this can be compensated for by optimizing the detectors to have a larger active volume.  

\begin{figure} [t]
   \centering
   %\begin{center}
   \begin{tabular}{c} 
   \includegraphics[height=6.2cm]{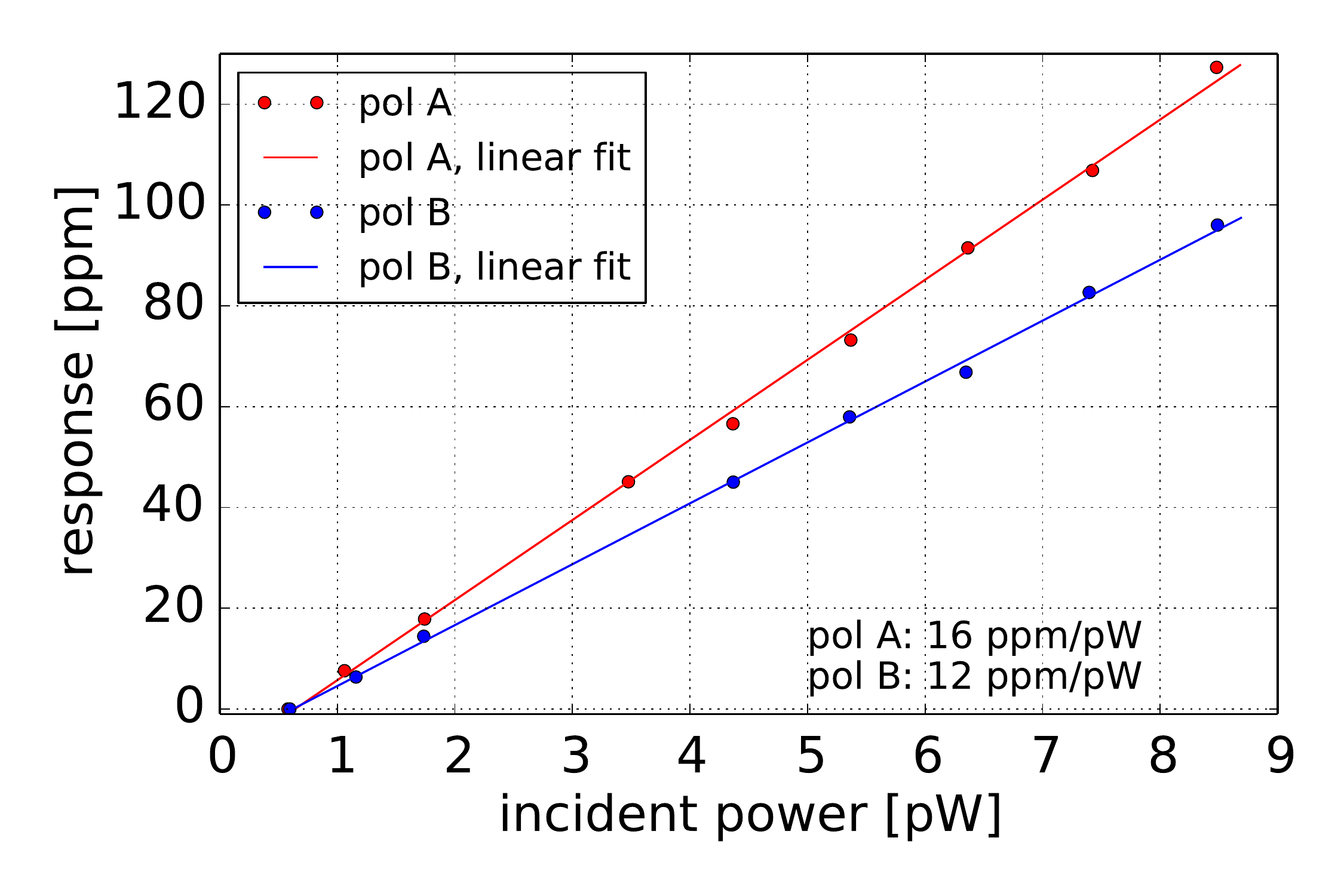}
   \includegraphics[height=6.1cm]{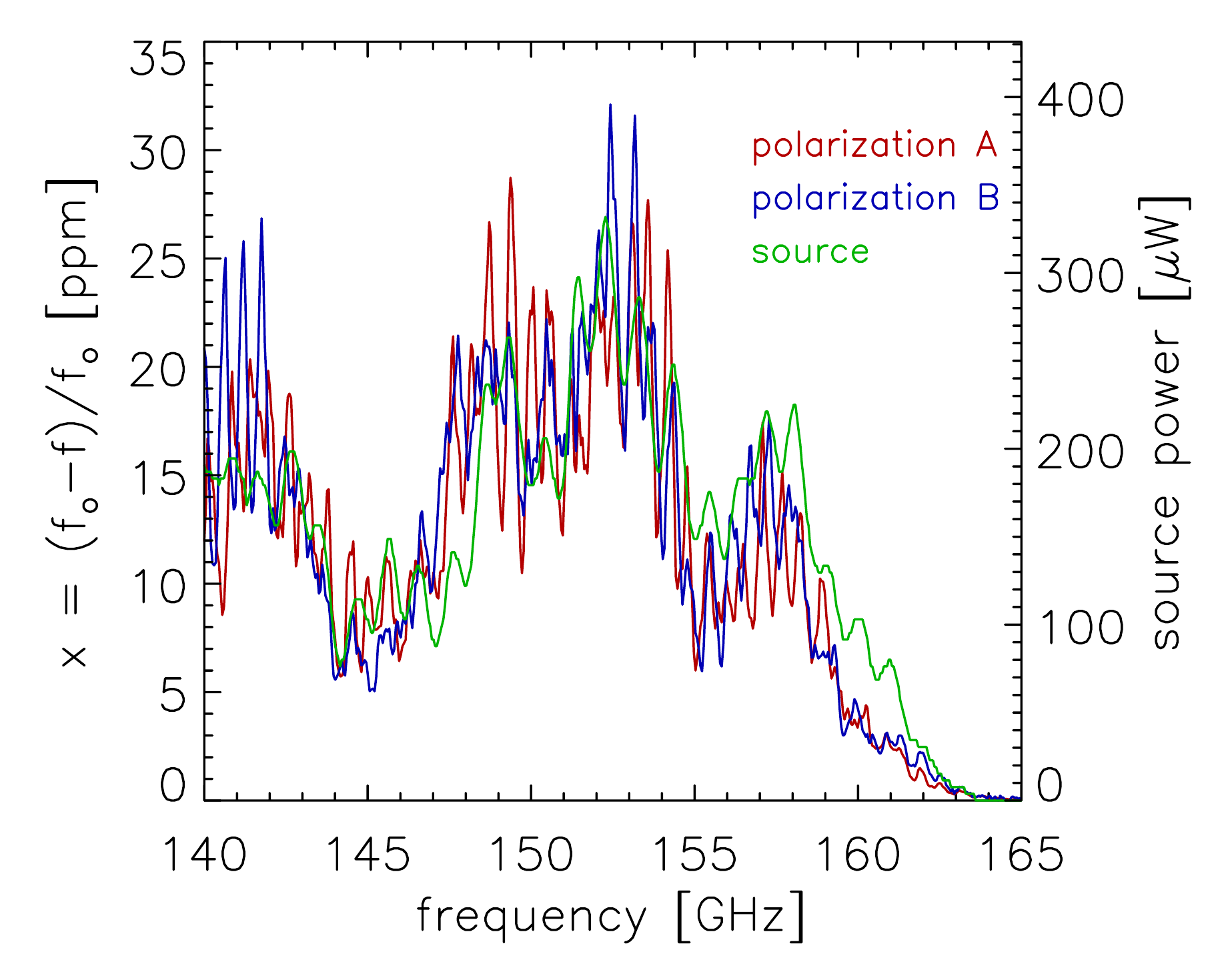}  
   \end{tabular}
   %\end{center} 
   \caption[example] 
   {\label{fig:response} 
   {\sl Left:} Plot showing the fractional frequency shift of both polarizations as a function of incident power. The slope of the line, or the responsivity, is 16~ppm/pW and 12~ppm/pW for the polarizations A and B respectively. This corresponds to 27 ppm/K and 22~ppm/K with a 4~K blackbody load, respectively. The conversion from temperature to incident power was calculated using a 127~-~170~GHz frequency band. 
	{\sl Right:} The fractional frequency response of both polarizations as a function of the frequency of the incident power. An electronic mm-wave source is swept from 140 to 165~GHz. The source power is plotted for reference in green, and there is good agreement between it and the fractional frequency shifts of the detectors.
    }
 \end{figure}

The optical responsivity was determined by measuring the fractional frequency shift of the resonator as a function of blackbody temperature. 
The fractional frequency shift is defined as $x$ = ($f_{0} - f$)/$f_{0}$, where $f$ and $f_{0}$ are the measured resonance frequency at a particular temperature and the maximum resonance frequency respectively. 
The blackbody temperature is converted to incident power by integrating from 127~to~170~GHz.
We expect the response of the detectors to follow a $\sqrt{P}$ dependence.
We instead see a linear response in both resonators as shown in Fig.~\ref{fig:response}, which has been observed previously\cite{mccarrick_2015,Hubmayr2014} and could be due to thermal quasiparticles or high-frequency leaks. 
Both resonator designs corresponding to the two polarizations have similar responses over the power range measured. 
We calculate a responsivity of 27~ppm/K and 22~ppm/K for the $A$ and $B$ polarizations, respectively, at 4~K. 
In terms of incident power, these responsivities correspond to 16~ppm/pW and 12~ppm/pW.

The spectrum of the devices was measured using a millimeter-wave source.
The source frequency is swept from 140~-~165~GHz and piped into the cryostat through a rectangular wave guide. 
The radiation is launched out of the waveguide oriented at 45$^{\circ}$ to the orthogonal resonators, so both devices should receive equal amounts of power.  
The plot of the fractional frequency response as a function of radiation frequency is shown in Fig.~\ref{fig:response}.
The source power is plotted for reference. 
The response of both detectors track each other well and also match the spectrum of the source itself.
The devices are currently undergoing FTS spectra measurements at Cardiff University. 

\begin{figure} [t]
 	\centering
%   \begin{center}
   \begin{tabular}{c} 
   \includegraphics[height=6.2cm]{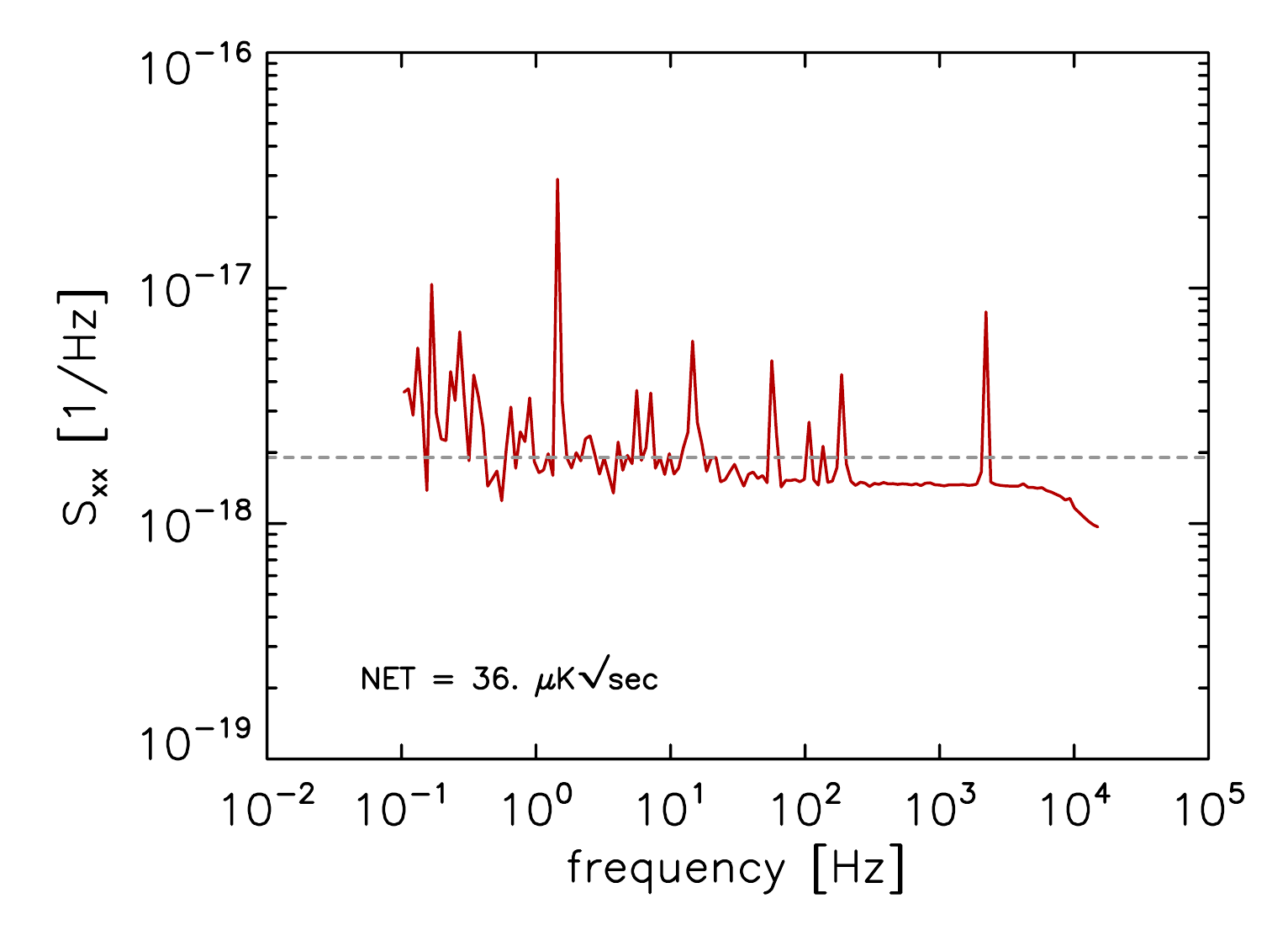}  
   \includegraphics[height=6.2cm]{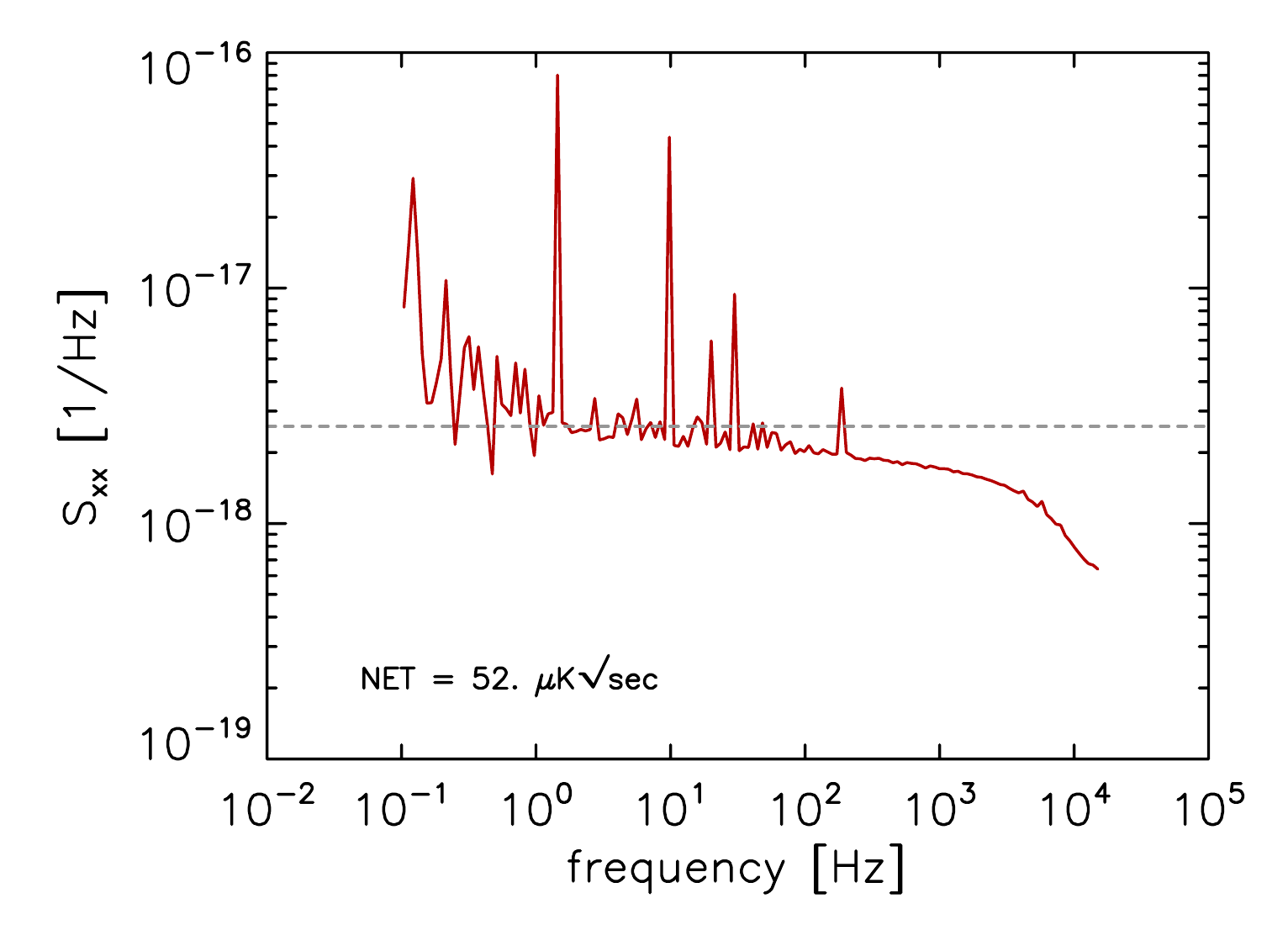}
   \end{tabular}
   %\end{center}
   \caption[example] 
   {\label{fig:noise} 
{\sl Left:} Representative noise spectra for a resonator of polarization A design.  The dotted line corresponds to a noise equivalent temperature (NET) of 36 $\mu$K$\sqrt{s}$ referenced to 4~K.
{\sl Right:} Representative noise spectra for a resonator of the orthogonal polarization B to that in the left plot. The dotted line corresponds to a NET of 52 $\mu$K$\sqrt{s}$ referenced to 4~K.
}
 \end{figure}

 %%%%%%%%%%%%%%%%%%%%%%%%%%%%%%%%%%%%%%%%%%%%%%%%%%%%%%%%%%%%%%%%%%%
Noise spectra of the detectors measured with the blackbody load temperature held at 3~K are shown in Fig.~\ref{fig:noise}. 
The spectra for the resonators of both devices is flat from approximately 1~Hz to $10^2$~Hz.
At low frequencies, fluctuations of the blackbody load temperature cause an increase in the noise. 
More work needs to be done to determine the precise detector 1/f knee frequency.
The amplifier noise is subdominant to the detector noise.
The noise equivalent temperature (NET) of the devices is found from the device noise level and the detector response. 
In the representative detectors shown in Fig.~\ref{fig:noise}, the NET is calculated to be 36~$\mu$K$\sqrt{s}$ and 52~$\mu$K$\sqrt{s}$ for polarizations A and B respectively. 
%

%%%%%%%%%%%%%%%%%%%%%%%%%%%%%%%%%%%%%%%%%%%%%%%%%%%%%%%%%%%%%%%%%%%%%%%%%%%%%%%%%%%%%%%%%%%%%%%
\section{Continued Work}
Future work will include measurements of polarization selectivity and beam-mapping of all detectors in the prototype array.
We will also fully investigate the noise sources of the devices.

The device design will be further optimized in future iterations. 
First, for ground-based observations, the expected optical-loading requires a greater absorbing volume.
We can increase the volume while maintaining the matched optical impedance by making small wiggles in the indcutor and increasing the film thickness. 
Second, we aim to reduce the cross-polarization response response by adding the chokes as originally designed.
Additionally, an adjustment to the absorber geometry directly under the horn aperture should decrease the cross-polarization further.  
As a demonstration of a LEKID array suitable for ground-based observations, we have fabricated a  hexagonal 271-element array based on the 20-element design discussed in this paper. 
The optimizations for ground-based devices discussed above have been incorporated. 
This array, to be read out with a single coaxial cable pair, will provide a demonstration of the high multiplexing factors achievable with LEKIDs. 
The corresponding horn array has also been fabricated. 
As all elements of the optical and readout systems exist, this array is in principle, deployable. 

%%%%%%%%%%%%%%%%%%%%%%%%%%%%%%%%%%%%%%%%%%%%%%%%%%%%%%%%%

\section{Conclusions}
We have successfully demonstrated dual-polarization devices for millimeter-wavelengths. 
The devices have flat noise within the device band and measured NETs of 36~and~52~$\mu K \sqrt{s}$ referenced to a 4~K load. 
The devices are responsive, with an optical responsivity of $\sim $20~ppm/K at a 4~K load. 
Initial tests show the two polarizations have similar responsivities.
We are currently performing measurements to determine the polarization selectivity.
Further optimizations to the detector design for ground-based observing have been implement in a 542-detector array that will be tested imminently. 
The large array will also allow the multiplexing capabilities to be further tested.
The design presented here and initial test results show that LEKIDs work well and are a promising technology for future CMB polarimetry experiments. 

%%%%%%%%%%%%%%%%%%%%%%%%%%%%%%%%%%%%%%%%%%%%%%%%%%%%%%%%%%

%%%%%%%%%%%%%%%%%%%%%%%%%%%%%%%%%%%%%%%%%%%%%%%%%%%%%%%

\acknowledgments % equivalent to \section*{ACKNOWLEDGMENTS}    
The devices were fabricated at JPL. 
We thank the Xilinx University Program for their  donation of FPGA hardware and software tools used in the readout system.
This research is supported in part by a grant from the Research Initiatives for Science and Engineering (RISE) program at Columbia University to B.R.J.
H. M. is supported by a NASA Earth and Space Sciences Fellowship.
%

% References
\bibliography{bjohnson_spie_2016} % bibliography data in report.bib
\bibliographystyle{spiebib} % makes bibtex use spiebib.bst

\end{document}